\documentclass[summary]{gass2017}
\makeatletter
\def\bstctlcite{\@ifnextchar[{\@bstctlcite}{\@bstctlcite[@auxout]}}
\def\@bstctlcite[#1]#2{\@bsphack
  \@for\@citeb:=#2\do{%
    \edef\@citeb{\expandafter\@firstofone\@citeb}%
    \if@filesw\immediate\write\csname #1\endcsname{\string\citation{\@citeb}}\fi}%
  \@esphack}
\makeatother
\usepackage[square,sort,numbers]{natbib}
\usepackage{url}
\title{Real-time searches for fast transients with Apertif and LOFAR}
\author{Yogesh Maan*\affref{ref1} and Joeri van Leeuwen\affref{ref1}\affref{ref2}}
\affiliation{
 \aff{ref1}{ASTRON, the Netherlands Institute for Radio Astronomy, Postbus 2, 7990 AA Dwingeloo, The Netherlands}
 \aff{ref2}{Astronomical Institute ``Anton Pannekoek'', University of Amsterdam, Science Park 904, 1098 XH Amsterdam, The Netherlands}
}

\begin{document}
\bstctlcite{IEEEexample:BSTcontrol}
\maketitle

\begin{abstract}
With the installation of a new phased array system called Apertif, the instantaneous
field of view of the Westerbork Synthesis Radio Telescope (WSRT) has increased to
8.7\,deg$^2$. This system has turned the WSRT in to an highly effective telescope to
conduct Fast Radio Burst (FRB) and pulsar surveys. To exploit this advantage, an
advanced and real-time backend, called the Apertif Radio Transient System (ARTS),
is being developed and commissioned at the WSRT. In addition to the real-time
detection of FRBs,
ARTS will localize the events to about 1/2600 of the field of view ---
essential information for identifying the nature of FRBs.
ARTS will also trigger real-time follow up with LOFAR of newly detected FRBs,
to achieve localization at arcsecond precision. We review 
 the upcoming time-domain surveys with Apertif, and present the current status
of the ongoing commissioning of the time domain capabilities of Apertif.
\end{abstract}

\section{Introduction}
The majority of searches for pulsars and fast transients have been
conducted using single dish radio telescopes. Despite providing
a poor handle on sky position of detected objects, single dish surveys
have been largely successful due to their capability of searching a
large sky volume for a given sensitivity and observing time. Pulsars
generally emit persistently, and are further localized in post-discovery follow-up.
However, Fast Radio Bursts (FRBs) --- bright pulses of a few millisecond duration
 --- are generally one-off events. Only one FRB is known to
repeat out of the 18 published discoveries to date \citep{Petroff16a,Spitler16}.
While it is nearly certain that FRBs
are generated at cosmological distances \citep[see, e.g.][]{Chatterjee17},
any further clue about their progenitors is awaited even after 10 years
of the first FRB discovery \citep{Lorimer07}. 
Probing the progenitors
and physical processes giving rise to such one-off events needs their
immediate follow-up at other parts of the spectrum, which in turn demands
well-constrained  positions. A survey that can couple discovery with immediate tight constraints
on the sky position is also extremely useful for
certain classes of pulsars, e.g., rotating radio transients (RRAT) and
intermittent pulsars. The upcoming pulsar and transient survey with the
Westerbork Synthesis Radio Telescope (WSRT) will combine the advantages
of a single dish and an interferometer: a large
field-of-view (FoV) as well as ability to tightly localize the source.
In the
following sections, we describe the unique features of the survey and
the instrumental developments that are in progress to realize these features.

\section{Exploring pulsars and FRBs with Apertif}
\begin{figure}[htbp]
  \centering
  \includegraphics[width=82mm]{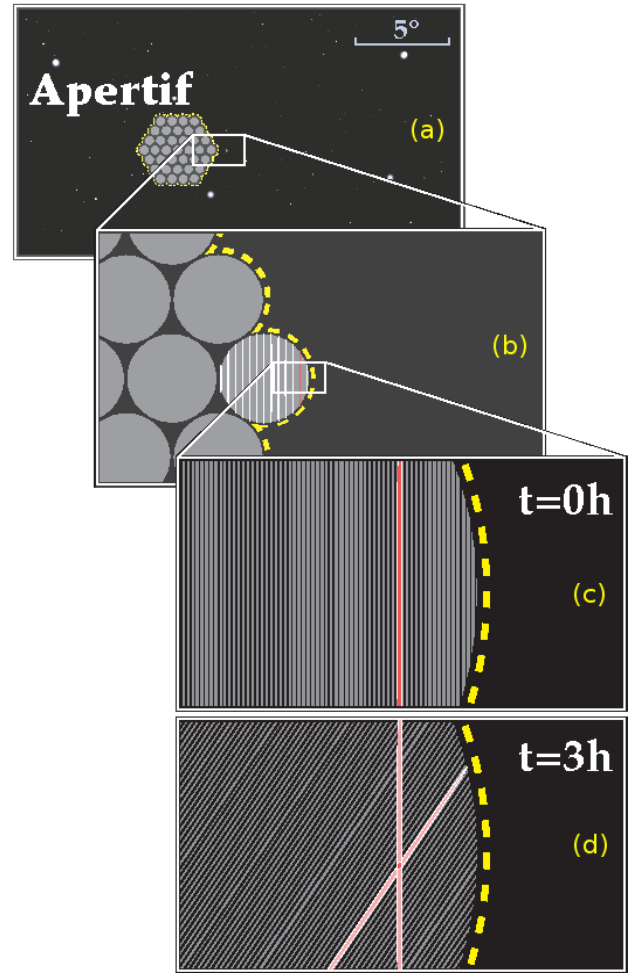}
  \caption{Panel (a) shows the 37 on-sky compound beams that Apertif facilitates
for each of the WSRT dishes. One of the tied-array beams and its grating lobes
are shown as vertical white lines within one compound beam in panel (b).
Synthesized beams --- position dependent combinations of various tied-array
beams --- are shown in panel (c). The red colored beam in this panel demonstrates
constraints on the position of an event detected in this beam. Panel (d) shows
the drastic improvement in localization from the second detection of a repeating
event. The position is constrained within the small, red circle at the
intersection of the two pink colored beams in this panel.}
  \label{arts_demo}
\end{figure}

A recently installed phased array system, the {APER}ture {T}ile
{I}n {F}ocus, or Apertif \citep{ovc+09}, facilitates 37 beams on the sky for 12 of
the WSRT dishes, and therefore increases the instantaneous FoV by nearly 37 times
(see panel (a) in Figure~\ref{arts_demo}) to about 8.7 deg$^2$. Apertif also doubles the
operating bandwidth to 300\,MHz, placed anywhere between 1100$-$1750\,MHz.
Such a large FoV and bandwidth, and the accumulated collecting area of several
25\,m diameter
dishes of the WSRT present an ideal combination for surveying the transient
radio sky. To exploit this combination, a new, advanced, real-time backend,
called the Apertif Radio Transient System (ARTS) is being developed and
commissioned at the WSRT. Within each of the 37 beams (hereonwards compound
beams or CB), ARTS will make several coherent or tied-array beams (TABs; also
see Section~\ref{sloc})
to retain the full sensitivity of the collecting area of 12 WSRT dishes.
ARTS further comprises of 40 computing nodes, each one consisting of dual 8-core
CPUs with 128 GB memory, 10\,GE cards, 2$-$6 GPUs providing a minimum performance
of 15\,Tflops/s and 30\,TB of disk for 8$-$12\,hr data caching.
Different nodes will process data from all the TABs within individual
CBs and search for FRBs and radio pulsars in \emph{real-time}. 
These technical advances 
enable a number of science cases, including a pulsar and FRB survey called ALERT, 
The Apertif LOFAR Exploration of the Radio Transient
Sky \citep[\url{www.alert.eu};][]{vanLeeuwen14}.
\begin{table*}
 \centering
  \caption{Comparison of ALERT's parameters with CHIME and Parkes surveys}
  \begin{tabular}{lclcc}
  \hline
            &  Field of View     & Localization            & S$_{\rm min}$ & FRB Detection Rate \\
            &  (sq. deg.)        & precision              & (mJy)         & (week$^{-1}$) \\
  \hline
  ALERT     &  8.7               & $\sim$arcsec            & 460           & $\sim$1 \\
            &                    & (ARTS: $25''\times15'$)&               &           \\
            &                    & (ARTS$_{\rm repeat}$: $25''\times25''$)&     &           \\
  Parkes    &  0.56              & $\sim14.1'$            & 180--250      & $0.05-0.5$ \\
  CHIME     &  134               & $\sim15'-30'$          & 160--270      & 20$-$380    \\
\hline
\end{tabular}
\newline
\newline
{See Section~2.1 for details on the conservative approach taken to estimate ALERT's
sensitivity and detection-rate, and Section~\ref{sloc} for those on stand-alone localization
by ARTS.
S$_{\rm min}$ is computed for 10$\sigma$ detection of a 5\,ms burst.
ALERT and Parkes surveys operate at L-band with similar bandwidths, while
CHIME will survey the sky between 400 and 800\,MHz. CHIME and Parkes survey parameters
are either directly noted or deduced from the information in:
\citep{BS14}, \citep{Newburgh14}, \citep{Connor16a}, \citep{Champion16}, \citep{Chawla17}.}
\end{table*}

ALERT primarily consists of 4 science cases. In the most compute intensive
science case, about 1-year of the WSRT time will be dedicated to survey the
whole northern sky with declination$\ge0^{\circ}$. FRBs will be detected in \emph{real-time}
and high time-resolution data for each detection will be saved for detailed
offline studies.
The detections will also trigger a \emph{real-time follow-up} of the bursts at
much lower frequencies using LOFAR, and quasi-real time follow-up at other energy
bands (e.g., X-rays, gamma-rays, etc.). In this dedicated survey, periodicity search
will be conducted using data cached in a 12-hr ring buffer. All the
time-domain survey data downsampled to 1\,ms, 1\,MHz and 1-bit will be archived.
ARTS will also conduct FRB and pulsar searches \emph{commensal}
with the Apertif imaging surveys \citep{aper16}. In these commensal searches, a newly developed beam-former operating in parallel to the correlator will form coherent TABs.  
As in the dedicated search, potential FRB
detections will trigger follow-up at other frequencies and energy bands,
and periodicity searches will be conducted using the data from the ring buffer.

The pulsar timing science case utilizes a single TAB in the central CB for
timing studies, and ARTS will facilitate a timing precision with systematic
instrumental errors less than 20\,ns. The pulsar data from the single TAB
are coherently dedispersed and folded in \emph{real-time}. In the fourth science
case, ARTS shall contribute to standard VLBI imaging, with a FoV of 0.000015\,deg$^2$
initially, based in a central TAB, and 0.25\,deg$^2$ eventually, when individual
dishes are streamed for VLBI. The following subsections provide a few salient
features of ALERT, primarily focusing on the dedicated search science case.

\subsection{Survey speed, sensitivity and FRB detection rate}
ARTS potentially facilitates full sensitivity of WSRT within Apertif's large FoV.
Assuming a 75\% aperture efficiency of 10 WSRT dishes, ARTS will detect a 5\,ms wide
320\,mJy burst at a 10$\sigma$ level.
If we also assume a \emph{TAB-forming efficiency}\footnote{This efficiency
factor also includes contribution from losses while forming various combinations
of TABs (see Section~\ref{sloc}). However, we expect it to be better than
the value assumed here.} of 0.7, a 460\,mJy burst will be detected at 10$\sigma$ significance.
Even with our conservative estimate of the sensitivity, more than 80\% of the FRBs
discovered so far\footnote{See FRB Catalogue at
\url{http://www.astronomy.swin.edu.au/pulsar/frbcat/}}
would be detectable, and most with high signal-to-noise ratio (S/N).
So far, majority of the FRBs have been discovered by Parkes telescope.
With nearly 16 times larger FoV, competing sensitivity and similar
operating frequency, ALERT will survey the sky an order of magnitude faster
than Parkes (see Table 1).

Constraints on the all-sky FRB rate are improving as the combined on-sky time
from various surveys increases. The recent all-sky rate of FRBs is estimated
to be $3.3^{+3.7}_{-2.2}\times 10^3$/day
above a fluence level of 3.8 Jy ms at 1.4 GHz \citep{Crawford16}. With this
rate and Apertif's large FoV, ALERT will detect 0.2$-$1.5 FRBs for every 24\,hrs
of observing time. This estimate is consistent with that predicted using
Monte Carlo simulations in \citep{Chawla17}; see their Figure 10. However,
the effects of TAB-forming efficiency parameter may not have been considered there.
Furthermore, the true underlying fluence and width distributions of FRBs are
still unknown and an excess of fainter, narrower FRBs can skew the detection~rate
estimates. So, with a conservative approach, we expect the detection rate to be
at the lower end of the above estimated range, i.e., about 1~burst/week.

Intermittent pulsars, RRATs, radio-transient magnetars and extremely nulling
pulsars are active only very sporadically --- some are on for less than 1\,s per day! Most of such objects are prone to be missed in typical short duration 
pulsar surveys. The long integration times in ALERT (typically 3$-$6\,hrs per pointing)
will be particularly sensitive to such neutron star types. This survey will
potentially double the number of currently known intermittent pulsars.

\subsection{Event localization by ARTS}\label{sloc}
Given the uniform spacing of the WSRT dishes, the TABs within individual CBs
have equally sensitive grating lobes. Hence each $\sim30'$ CB can be
covered by only about 12 TABs and their grating lobes (see panel (b) in
Figure~\ref{arts_demo}).
However, the grating lobe separations are frequency dependent, and the
difference in these separations for beams at the twos edges of ALERT's
large fractional bandwidth is quite significant. FRBs are broadband
events, and to integrate any weak signals over the entire band, ARTS
combines various TABs such that the complete telescope response for a
certain location on the sky is recovered. This results in \emph{synthesized
beams (SB)}, and by positioning these SBs at the locations of the highest-frequency
TABs, each compound beam is covered by 71 SBs (see panel (c) in Figure~\ref{arts_demo}).
Hence, Apertif's full FoV is covered by over 2600 (71$\times$37) coherent SBs, and
data from each of the SBs will be searched for transients in real-time.

The TABs also posed a degeneracy in the sky position of an event to any of
their grating lobes. The synthesized beams break this degeneracy, and
the sky position of a single event could be constrained to one SB, i.e.,
within $25''\times15'$.
As the earth rotates, the position of the SBs will change with time. This
aspect implies that the sky position of an event could be constrained to
$25''\times25''$, even with just one repeat detection of the same event
(see panel (d) in Figure~\ref{arts_demo}).
Hence, ARTS will provide discovery positions of \emph{repeating}~FRBs, RRATS,
and all the new pulsars constrained within $25''\times25''$.

\subsection{ALERT triggers: Immediate follow-ups and \emph{arcsecond} localization}
The frequency-dependent propagation delays and the cosmological origins
of FRBs imply their arrival at low frequencies (e.g. LOFAR's 150\,MHz)
would be delayed by several minutes compared to that at 1.4\,GHz.
ARTS will exploit this aspect to trigger immediate LOFAR follow-ups of each
of the FRBs that it will detect in real-time. A detection in one of the
LOFAR TABs will provide (1) an unprecedented FRB detection at
such low frequencies, (2) crucial information about the spectral behavior
and hence the underlying emission mechanism, and (3) constraints on the
sky position to arcsecond precision. The accurate positions thus obtained
will be used to trigger optical and high energy follow-ups, to locate
faint host galaxies and detect possible afterglows, among other probes.

With these unique features, ALERT will facilitate FRB population studies
by detecting several tens to hundreds of new bursts, studying their radio spectra, and
enabling a detections  at optical and high energy bands by localizing
with arcsecond precision.
\begin{figure}[htbp]
  \centering
  \includegraphics[width=75mm]{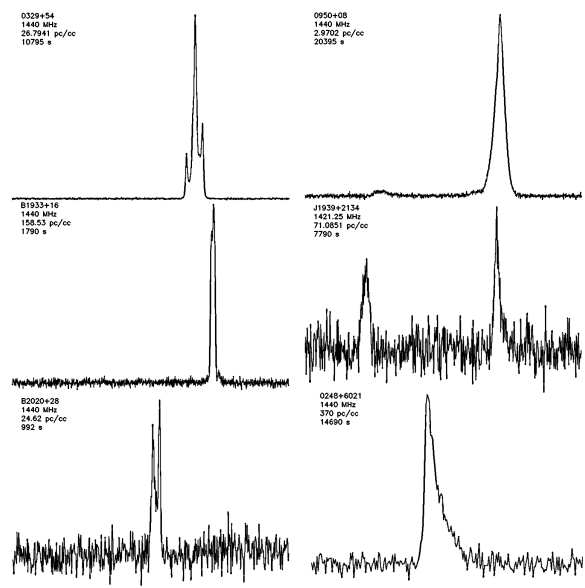}
  \caption{Average profiles of several pulsars detected in commissioning observations.}
  \label{comm_plot}
\end{figure}

\section{ARTS development and commissioning: Current status}
The first hardware component of ARTS enables the pulsar timing science case,
and is currently (Feb. 2017) being commissioned. The high precision pulsar timing pipeline
is implemented on a powerful machine consisting of two CPUs with 12 cores each,
128\,GB of memory, with a dual Titan-X (Maxwell) GPU, and more than 30\,TB of disk space to store the
pulsar timing observations. Data from a single TAB are coherently dedispersed
and folded over the pulsar period in \emph{real-time}. Folded profiles with 384
channels across 300\,MHz bandwidth are written out every 10\,s for offline
processing and timing studies. The instrument saw the first light from the
bright pulsar {B0329+54} in August 2016 using only 1 polarization and a
small bandwidth of 18.75\,MHz. Since then, extensive commission has been
underway testing and ironing out wrinkles at various processing steps and making
steady progress. We have now progressed to the timing pipeline operating
flawlessly on the dual-polarization, full bandwidth (300\,MHz) data streams.

Using a single element of the phased-array feed and only one WSRT dish, we
have successfully detected several pulsars during commissioning. These detections
sample a variety of pulsar parameters, viz. faint to strong (5$-$200\,mJy), fast to
slow (1.5$-$720\,ms), and low to high dispersion measure (3$-$370\,pc\;cm$^{-3}$).
Average profiles obtained during commissioning observations of
some of these pulsars are shown in Figure~\ref{comm_plot}.

The software pipeline to be used in dedicated as well as commensal FRB searches
has been completed and successfully tested on archive data of a burst discovered by the
Parkes telescope. The computing performance of the pipeline to carry out real-time searches
on more than 2600 SBs has been achieved, and was benchmarked for various GPUs. The complete GPU back-end cluster 
arrives in 2017Q3. We expect to complete the commissioning,
and start the dedicated observations and detecting FRBs towards the end of 2017.

\section{Summary}
The ALERT real-time searches through over 2600 full-sensitivity synthesized beams
within Apertif's large FoV will potentially discover several 10s to 100s of new
FRBs. The arcsecond precision localization of the bursts by ARTS-triggered
LOFAR follow up is a unique feature of ALERT, and such a precision seems to
remain unsurpassed by any of the current or upcoming surveys in next few years.
This much needed
localization will enable the follow-ups in optical, X-ray and gamma-ray wavelengths
as well as in other radio frequency bands to identify the host galaxies, detect any
possible afterglows in high energy bands and understand the progenitors of these
mysterious bursts.

\section{Acknowledgements}
The development and commissioning of ARTS is carried out
by a large team of engineers and astronomers, including active
participation from the presenting author (YM) and PI (JvL).
The research leading to these results has received funding from the European Research Council
under the European Union's Seventh Framework Programme (FP/2007-2013) / ERC Grant 
Agreement n. 617199.


\end{document}